\documentclass [a4paper]{article}

\usepackage{psfig,pstricks,multicol,pst-grad,color,fancybox,graphics}
\usepackage{latexsym}
\usepackage{amssymb}
\usepackage{enumerate}
\usepackage{bbm}

\newcommand{\Bra}[1]{\ensuremath{\langle#1|}}
\newcommand{\Ket}[1]{\ensuremath{|#1\rangle}}
\newcommand{\BraKet}[2]{\ensuremath{\langle #1|#2\rangle}}
\newcommand{\KetBra}[1]{\ensuremath{| #1 \rangle \langle #1 |}}

\newcommand{\Eins}{\ensuremath{\mathbbm 1}}
\newcommand{\HH}{\ensuremath{\mathcal{H}}}

\newcommand{\BE}{\begin{equation}}
\newcommand{\EE}{\end{equation}}
\newcommand{\kommentar}[1]{}

\begin{document} 
\title{Investigating three qubit entanglement with local \\
measurements} 
\author{Otfried G\"uhne$^1$ and Philipp Hyllus \\[5mm]
Institut f\"ur Theoretische Physik, Universit\"at Hannover,\\
 Appelstra\ss e 2, 30167 Hannover, Germany.
}

\maketitle    

\footnotetext[1]{To whom correspondence should be adressed. 
E-mail: guehne@itp.uni-hannover.de}

\noindent
{\bf Abstract:}
In this paper we describe how three qubit entanglement can 
be analyzed with local measurements. For this purpose we 
decompose entanglement witnesses into operators which
can be measured locally. Our decompositions are optimized
in the number of measurement settings needed for  
the measurement of one witness. Our method allows to detect 
true threepartite entanglement and especially GHZ-states 
with only four measurement settings. 


\section{Introduction}

Entanglement is one of the most puzzling features of quantum theory 
and of great importance for quantum information theory. It is the 
resource that makes various quantum protocols possible 
that perform certain tasks better than it would 
be possible with purely classical methods \cite{2000:chiara}. 
The investigation and characterization of entanglement with 
experimental and theoretical methods itself is therefore a task 
of great importance. 

From an experimental point of view it is important to find effective  
techniques for the production and detection of entanglement. For the 
detection of entanglement several strategies are known: One can 
determine the whole density matrix \cite{2002:Thew} and then try to apply 
one of the necessary or sufficient entanglement criteria, e.g. the
PPT criterion \cite{1996:peres,1996:horod}. One can also look for a 
violation of Bell inequalities \cite{2001:WWreview}. Furthermore,
there have been several proposals of detecting entanglement 
without estimating the whole density matrix \cite{2002:phorod,2001:phorod}.

However, all these nice ideas have also some disadvantages. Estimating the 
whole density matrix requires a lot of measurements, and in fact one does
not need to know the whole matrix in order to check whether it is entangled 
or not. On the other hand, looking for a violation of Bell inequalities 
may not suffice for a making a decision, since there are entangled 
states which do not violate any known Bell inequality 
\cite{2000:WW,2001:WW,2001:Zukowski}. It has even been 
conjectured that entangled states with a positive partial transpose do 
not violate any Bell inequality at all \cite{1999:peres}. Finally, the 
recent proposals of detection require collective measurements on several 
qubits which makes them hard to implement with the present techniques.

Furthermore, the schemes mentioned above are in some sense 
too general for many experimental situations. They assume that no 
{\it a priori} information about the state is given. However, 
in a typical experimental situation one often tries to prepare a 
certain pure state. Due to imperfections of the experimental 
apparatus the output state will be a mixture of the desired state
and some noise instead.
In this case it is desirable to know whether the produced state 
is still entangled or not.

In \cite{2002:guehne1,2002:guehne2} we proposed a general scheme for 
entanglement detection for the case that some knowledge about the 
state is given. In this scheme we only want to use local projective 
measurements, since these measurements can easily be implemented 
in a lab. In addition, we would like to decrease the number of 
measurements needed to the minimum, of course.

The scheme relies on the well known concept of witness operators 
\cite{1996:horod,2000:terhal1}. An hermitean operator 
$W$ is called a witness operator 
(or entanglement witness) detecting the entangled state $\varrho_e$ 
if $Tr(W\varrho_e)<0$ while $Tr(W\varrho_s)\geq 0$ holds for 
all separable states $\varrho_s.$ Thus if a measurement 
yields $Tr(W\varrho)<0$ then the state $\varrho$ is entangled with 
certainty. As a consequence of the Hahn-Banach theorem it 
follows that for every entangled state $\varrho_e$ there 
exists such an entanglement witness and for many states it is known
how to construct such witnesses \cite{2000:lewenstein}. 
After having constructed the witness, we decompose it into a sum of 
local measurements, then the expectation value can be measured with 
simple methods. This decomposition has to be optimized in a certain 
way since we want to use the smallest number of measurements possible. 

Our paper is organized as follows: 
The first section deals with the decomposition into local 
measurements because this is the core of our approach. We 
define there what we understand by an optimized decomposition. We 
would like to remark here that finding optimized decompositions is 
in general a hard task, much harder than constructing entanglement 
witnesses. In the second section we illustrate our approach with 
an example of a two qubit system. We construct the witness and 
determine its optimal decomposition. Finally in the third and main 
part we apply this idea to three qubit systems. 
We explain how GHZ-type and W-type entanglement can be detected
with local measurements. We also determine the minimal number of 
measurements needed for this.  

\section{Local decompositions}
Assume that we have an hermitean operator $H$ acting on a tensor product
$\HH=\HH_A\otimes\HH_B\otimes ... \otimes \HH_Z$ of two or more 
(but finite) dimensional Hilbert 
spaces. In order to slenderize the notation we look here at the case that 
we have a bipartite $N \times N$-system: $\HH=\HH_A\otimes\HH_B$ 
with $\mbox{dim}(\HH_A)=\mbox{dim}(\HH_B)=N.$  But all definitions 
in this section can be extended to more parties in an obvious manner. 
In order to measure the expectation value of this operator locally, 
we have to decompose it in a sum of tensor products of operators acting 
on only one system, or, equivalently, we have to decompose it into a sum
of projectors onto product vectors:
\begin{equation}
H = \sum_{i=1}^{m} A_i\otimes B_i = 
\sum_{i=1}^{n} c_{i}\KetBra{e_{i}} \otimes \KetBra{f_{i}}. 
\label{pvdecomposition}
\end{equation}
By measuring the expectation value of the projectors 
$\KetBra{e_{i}} \otimes \KetBra{f_{i}}$ and adding the results 
with the weights $c_i$ this decomposition (\ref{pvdecomposition}) 
can be measured locally.  
There are, of course, many possibilities of finding a decomposition
like (\ref{pvdecomposition}). So we have to optimize the decomposition 
in a certain sense. But there are even several possibilities of defining 
an optimized decomposition. 

On the first sight one might try to minimize the number of 
product vectors corresponding to minimizing $n$ in (\ref{pvdecomposition}). 
This optimization procedure is already known from the literature, 
it was considered in \cite{1998:sanpera}. There it was shown that 
for a general operator acting on two qubits one needs
five product vectors and also a constructive way of computing these 
product vectors was given.

However, since we want to construct an experimentally accessible 
scheme for entanglement detection it is natural to look for a 
decomposition where Alice and Bob have to perform the smallest number of 
measurements possible. 
By ``measurements'' we understand here von Neumann (or projective) 
measurements, since they can be easily implemented. Such a measurement
for Alice corresponds to a choice of an orthonormal basis of $\HH_A,$
and Bob has to choose an orthonormal basis $\HH_B$, too. So any operator of 
the form   
\begin{equation}
M = \sum_{k,l=1}^{N}c_{kl}\KetBra{e_k}\otimes\KetBra{f_l}
\label{lvnmdefinition}
\end{equation}
with $\BraKet{e_s}{e_t}=\BraKet{f_s}{f_t}=\delta_{st}$ can be 
measured with only one collective setting of measurement devices 
of Alice and Bob. Alice and Bob can distinguish the 
states $\Ket{e_k f_l},$ measure the probabilities of these states 
and add their results with the weights $c_{kl}$ to measure $M.$
We call an operator which can be measured with one measurement 
setting (like $M$ in Eq.~(\ref{lvnmdefinition})) a \emph{local von 
Neumann measurement} (LvNM).

Having understood what can be realized with one measurement 
setting, we can state another optimization strategy. We want to find
a decomposition of the form 
\begin{equation}
H=\sum_{i=1}^{k}\sum_{k,l=1}^{N}c^{i}_{kl}\KetBra{e^{i}_{k}}
\otimes\KetBra{f^{i}_{l}}
\label{problemdefinition}
\end{equation}
with $\BraKet{e^i_s}{e^i_t}=\BraKet{f^i_s}{f^i_t}=\delta_{st}$ and
a minimal $k.$  This $k$ is the number of collective measurement 
settings Alice and Bob have to use in order to measure $H.$ This 
optimization strategy is the aim we are considering in this paper
when we talk about ``optimized'' decompositions. 

The reader should note that minimizing $m$ in (\ref{pvdecomposition})
is not the same as our optimization strategy. 
However, for systems of dimension $N$ greater than 2 it might be 
more suitable to decompose the witness as 
$W=\sum_{i}^{m}A_{i}\otimes B_{i}$. As shown in \cite{2002:phorod2},
the expectation values of operators $A_{i}$ or $B_{i}$ can
be measured by a POVM with a single qubit as ancilla 
instead of counting clicks for all possible N outcomes of the 
operator. Also minimizing the number
of product vectors ({\it i.e.} minimizing $n$ in (\ref{pvdecomposition}))
is not the same. This will become clear in a few seconds, when we study 
two qubits. 

\section{Two qubits}
We illustrate the method by considering an experiment that aims at 
producing a certain 2-qubit state $\Ket{\Psi}=a\Ket{01}+b\Ket{10}$ 
written in the Schmidt decomposition, {\it i.e.} $a,b\ge 0, a^{2}+b^{2}=1.$ 
Due to imperfections, the produced state will rather be
\begin{equation}
    \varrho_{p,d}=p\KetBra{\Psi}+(1-p)\sigma,
\end{equation}
where we assume that the noise state $\sigma$ 
lies inside a separable ball of radius $d$ around the totally 
mixed state, {\em i.e.} $\|\sigma-\Eins/4 \|\le d$ for some
norm. Our aim is to provide a local experimental method to tell 
whether the produced state is entangled or not, based 
on witness operators.

Since we only want to explain our basic idea, we assume here that 
we have white noise, this means $d=0.$ The case $d>0$ is studied 
in greater detail in \cite{2002:guehne2}.
We would like to stress that our assumption $d=0$ is in some sense
artificial. By this we mean that if $d=0$ there is a simple way of determining 
whether $\varrho_{p,0}$ is entangled or not: One can just measure 
\emph{any} operator $A$ (which fulfills $Tr(A)\neq 4Tr(A\KetBra{\psi})$)
and compute $p$ from the expectation value of this operator. With this
information the density matrix $\varrho_{p,0}$ can be constructed and 
the PPT criterion can be used to decide whether it is entangled or 
not. This is not possible for obvious reasons if $d>0$.

To reach our goal we first have to construct a proper entanglement witness.
For NPPT entangled states, {\it i.e.} entangled states with a non-positive
partial transpose, the construction of a witness is relatively easy:
The partial transpose of a projector onto an eigenvector of 
$\varrho^{T_{B}}$ with negative eigenvalue does the job. 
Here $T_{B}$ denotes the partial transposition with respect 
to subsystem $B.$

The state $\varrho_{p,0}$ has one possibly negative eigenvalue, 
namely ${\lambda_-}=(1-p)/4-abp$, with the corresponding eigenvector
\begin{equation}
  \Ket{\psi_-}=\frac{1}{\sqrt{2}}(\Ket{00}-\Ket{11})
\end{equation}
that is independent of $a$ and $p$. Then 
$W_{0}=\KetBra{\psi_-}^{T_{B}}$ is an entanglement witness detecting
$\varrho_{p,0}$ since 
\begin{equation}
 Tr(\KetBra{\psi_-}^{T_{B}})\varrho_{p,0}
 =Tr\KetBra{\psi_-}(\varrho_{p,0}^{T_{B}})={\lambda_-} <0.
 \label{eq:goodwitness}
\end{equation}

Having constructed the witness, all that remains is to decompose it.
Defining $\Ket{z^{\pm}}=\Ket{0,1}$, 
$\Ket{x^{\pm}}=(\Ket{0}\pm\Ket{1})/\sqrt{2}$ 
and $\Ket{y^{\pm}}=(\Ket{0}\pm i\Ket{1})/\sqrt{2}$ we can decompose
the more general
witness $\KetBra{\phi}^{T_{B}}$ for $\Ket{\phi}=\alpha\Ket{00}+\beta\Ket{11}$
as follows:
\begin{eqnarray}
  \label{eq:qubitwitness}
  {\KetBra{\phi}^{T_B}}
  &=& \alpha^2\KetBra{00}+\beta^2\KetBra{11}+
    \alpha\beta(\Ket{01}\Bra{10}+\Ket{10}\Bra{01})
      \nonumber\\
  &=& \alpha^2\KetBra{z^+ z^+}+\beta^2\KetBra{z^- z^-}+
  \alpha\beta\left(\KetBra{x^+ x^+}+\right. 
  \nonumber\\
  & & \left.+\KetBra{x^- x^-}-\KetBra{y^+ y^-}-\KetBra{y^- y^+}\right)
  \nonumber\\
  &=&\frac{1}{4}\left(\Eins \otimes \Eins +\sigma_z\otimes \sigma_z 
  +(\alpha^2-\beta^2)(\sigma_z\otimes\Eins+\Eins\otimes\sigma_z) \right. 
  \nonumber \\
  & &\left. +2\alpha\beta(\sigma_x\otimes \sigma_x+\sigma_y\otimes 
  \sigma_y)\right),
\label{anton}
\end{eqnarray}
where the $\sigma_{i}$ are the Pauli matrices. This way of decomposing 
a witness can be used for higher dimensions and for systems of more 
than 2 parties by using
\begin{eqnarray}
\Ket{01}\Bra{10}+\Ket{10}\Bra{01}
&=& \KetBra{x^+ x^+}+\KetBra{x^- x^-}-\nonumber \\
& & -\KetBra{y^+ y^-}-\KetBra{y^- y^+}.
\end{eqnarray}
The local measurement of the general witness of Eq.~(\ref{eq:qubitwitness}) 
requires three measurements of Alice and of Bob in the $x,y$ and $z$ 
direction. This is also true for the special case 
$\alpha=-\beta=1/\sqrt{2}$ corresponding to $W_{0}$. 
It is not possible to evaluate the witness with less than 
three LvNMs: 
\\
\\
{\bf Proposition 1.} The witness $W_0$ can not be decomposed into 
less than three LvNMs, therefore the decomposition (\ref{anton})
is optimal.
\\
\emph{Proof.}
The proof was first given in \cite{2002:guehne1}, we repeat
it here because we extend it later to three qubit systems.
Consider a decomposition requiring two measurements:
\begin{equation}
{\KetBra{\psi}}^{T_B}=
\sum_{i,j=0}^1 c^1_{ij} \KetBra{A^1_{i}} \otimes \KetBra{B^1_{j}}+
\sum_{i,j=0}^1 c^2_{ij} \KetBra{A^2_{i}} \otimes \KetBra{B^2_{j}}, 
\label{2x2decomposition}
\end{equation} 
where $\{\Ket{A_{i}}\}$ and $\{\Ket{B_{i}}\}$ are
orthonormal bases for $\HH_A$ and $\HH_B$, respectively.
With the help of a Schmidt decomposition as above we can write 
${\KetBra{\psi}}^{T_B}=\sum_{i,j=0}^{3} \lambda_{ij} \; 
\sigma_i \otimes \sigma_j$ with  
\begin{equation}
(\lambda_{ij})=
\left( \begin{array}{cccc}
\frac{1}{4}       &0            &0 & \frac{\alpha^2-\beta^2}{4}\\
0                 &\frac{\alpha\beta}{2} &0 & 0                \\
0                 &0            &\frac{\alpha\beta}{2} & 0     \\
\frac{\alpha^2-\beta^2}{4} &0            &0     & \frac{1}{4} 
\end{array}\right). 
\label{lambdamatrix}
\end{equation}
Note that the $3 \times 3$ sub-matrix in the right bottom corner 
is of rank 3. Now we write any projector on the rhs
of (\ref{2x2decomposition}) 
as a vector in the Bloch sphere: 
$\KetBra{A^1_{0}}=\sum_{i=0}^3 s^A_i \sigma_i$ is represented by
the vector $s^{A^1_0}=(1/2,s^A_1, s^A_2,s^A_3)$ and 
$\KetBra{A^1_1}$ by $s^{A^1_1}=(1/2,-s^A_1,-s^A_2,-s^A_3);$
$\KetBra{B^1_{0}}$ can be written similarly. If we expand the 
first sum on the rhs of (\ref{2x2decomposition}) in the 
($\sigma_i \otimes \sigma_j$) basis, the $ 3 \times 3$ sub-matrix 
in the right bottom corner is given by 
$(c^1_{00}-c^1_{01}-c^1_{10}+c^1_{11})
(s^A_1,s^A_2,s^A_3)^T(s^B_1,s^B_2,s^B_3).$ This matrix is of 
rank one.
The corresponding sub-matrix from the second LvNM on the rhs of 
(\ref{2x2decomposition}) is also of rank one and we arrive at a 
contradiction: No matrix of rank three can be written as a sum of two 
matrices of rank one.$\hfill\Box$

Please note that the decomposition~(\ref{anton}) requires 6 projectors 
onto product vectors (PPV). By applying the method of 
\cite{1998:sanpera} it is possible to decompose the witness using 
only 5 PPV:
\begin{equation}
  \KetBra{\psi}^{T_B}=\frac{(\alpha+\beta)^2}{3}\sum_{i=1}^3 
  \KetBra{ A'_i  A'_i} - \alpha \beta (\KetBra{01}+\KetBra{10}),
\end{equation}
where we have used the definitions 
\begin{eqnarray}
  \Ket{A'_1}& = & e^{i\frac{\pi}{3}} \cos(\theta) \Ket{0} 
  + e^{-i\frac{\pi}{3}} \sin(\theta)\Ket{1} \\
  \Ket{A'_2}& = & e^{-i\frac{\pi}{3}} \cos(\theta) \Ket{0} 
  +e^{i\frac{\pi}{3}} \sin(\theta)\Ket{1} \nonumber \\ 
  \Ket{A'_3}& = & \Ket{A'_1}+\Ket{A'_2} \nonumber \\
  \cos(\theta)&=&\sqrt{\alpha/(\alpha+\beta)} \nonumber \\
    \sin(\theta)&=&\sqrt{\beta/(\alpha+\beta)}. 
\end{eqnarray}
However, with this decomposition the measurement of the witness
would require 4 local correlated measurement settings, hence the 
two optimization strategies are really different.

\section{Three qubits}
The state space for three qubits has a much richer structure concerning
entanglement than the space of two qubits. Let us briefly remind the reader
of some well known facts about three qubits. We first consider pure 
states. There are two classes of states which are not genuine 
threepartite entangled: The fully separable states, which can be 
written as
\begin{equation}
\Ket{\phi_{fs}}_{ABC}=
\Ket{\alpha}_A\otimes\Ket{\beta}_B\otimes\Ket{\gamma}_C,
\end{equation}
and the biseparable states which can be written as
a product state in the bipartite system, which is created, if two of 
the three qubits are grouped together to one party. One example is
\begin{equation}
\Ket{\phi_{bs}}_{A-BC}=
\Ket{\alpha}_A\otimes\Ket{\delta}_{BC}.
\end{equation} 
There are three possibilities of grouping two qubits together, 
hence there are three classes of biseparable states. The genuine threepartite
entangled states are the states which are neither fully separable nor 
biseparable.
Given two threepartite states, $\Ket{\phi}$ and $\Ket{\psi},$ one can 
ask whether it is possible to transform  $\Ket{\phi}$ into $\Ket{\psi}$
with local operations and classical communication, without requiring that
this can be done with probability 1. These operations are called 
stochastic local operations and classical communication (SLOCC). It turns
out \cite{2000:duer} that $\Ket{\phi}$ can be transformed 
into $\Ket{\psi}$ iff there exist operators $A,B,C,$ acting on the space 
of one qubit with
\begin{equation}
\Ket{\psi}=A \otimes B \otimes C \Ket{\phi}.
\end{equation}
Surprisingly, it was proven in \cite{2000:duer} that there are two classes of 
genuine threepartite entangled states which cannot be transformed
into another by SLOCC. One class, the class of GHZ-states can be 
transformed by SLOCC into 
\begin{equation}
\Ket{GHZ}=1/\sqrt{2}(\Ket{000}+\Ket{111}),
\end{equation}
the other class, the class of W-states can be converted into
\begin{equation}
\Ket{W}=1/\sqrt{3}(\Ket{100}+\Ket{010}+\Ket{001}.
\end{equation}

Now we can classify the mixed states according to \cite{2001:acin}.
We define a mixed state $\varrho$ as fully separable if $\varrho$
can be written as a convex combination of fully separable pure states. 
A state $\varrho$ which is not fully separable is called biseparable 
if it can be written as a convex combination of biseparable pure states. 
One can, of course, define three classes of biseparable mixed states 
with respect of one of the three partitions as well. 
Finally, $\varrho$ is fully entangled if it is neither biseparable nor
fully separable. There are again two classes of fully entangled mixed states,
the W-class and the GHZ-class.  $\varrho$ belongs to the W-class, if it can 
be written as a convex combination of pure W- states, and to the GHZ-class
otherwise. 
Taking into account that the set of all states is also a convex set, 
one obtains an ``onion''-structure. This 
structure is shown in Fig. 1.

\begin{figure}[h!!]
\centerline{\psfig{figure=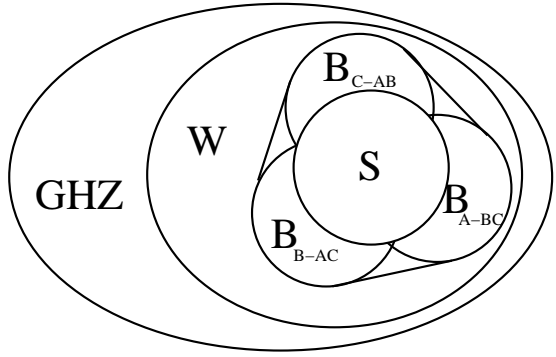,width=0.6\textwidth}}
\caption{The structure of the set of three qubit states: 
They can be (S)eparable, (B)iseparable, (W)-entangled, or 
(GHZ)-entangled.}
\end{figure}
In the same reference also witnesses for the detection of GHZ-type 
and W-type states have been constructed. Here a want to compute the 
optimized decompositions of these operators. 

For the GHZ-class a witness operator is given by
\begin{equation}
W_{GHZ}=\frac{3}{4}\Eins-\KetBra{GHZ}.
\label{ghzwitnessdefinition}
\end{equation}
If $\varrho$ is a mixed state with $Tr(\varrho W_{GHZ})<0$ the 
state $\varrho$ belongs to the GHZ-class. A decomposition of 
$W_{GHZ}$ can be computed with similar methods as for the two
qubit case, it yields
\begin{eqnarray}
W_{GHZ}&=&
\frac{1}{8}\Bigl(
5 \cdot\Eins\otimes\Eins\otimes\Eins-
\Eins\otimes\sigma_z\otimes\sigma_z-
\sigma_z \otimes \Eins \otimes \sigma_z- 
\sigma_z \otimes \sigma_z \otimes \Eins -  
\nonumber \\
& & 
- 2 \cdot \sigma_x \otimes \sigma_x \otimes \sigma_x +
1/2 \cdot 
(\sigma_x + \sigma_y) \otimes(\sigma_x + \sigma_y) 
\otimes(\sigma_x + \sigma_y)+
\nonumber \\
&& 
+ 1/2 \cdot 
(\sigma_x - \sigma_y) \otimes(\sigma_x - \sigma_y) 
\otimes(\sigma_x - \sigma_y) 
\Bigr).
\label{ghzwitnesszerlegung}
\end{eqnarray}
This witness can be measured with four collective measurement settings. 
Now we have to show that this decomposition is optimal.
\\
\\
{\bf Proposition 2.}
The witness (\ref{ghzwitnessdefinition}) can not be measured with three
LvNMs, {\it i.e.} the decomposition (\ref{ghzwitnesszerlegung}) is optimal.
\\
\emph{Proof.} The proof is an extension of the two qubit case. First, 
we write the witness in the 
$\sigma_{i} \otimes \sigma_{j} \otimes \sigma_{k}$ basis:
$ W_{GHZ}= 1/8 \sum_{i,j,k=0}^{3}\lambda_{ijk}
\sigma_{i} \otimes \sigma_{j} \otimes \sigma_{k},$ 
and from (\ref{ghzwitnesszerlegung}) we obtain:
\begin{eqnarray}
\lambda_{ij0} 
&=&
\left( \begin{array}{cccc}
5&0&0&0\\
0&0&0&0\\
0&0&0&0\\
0&0&0&-1 
\end{array}\right) =: A^{(0)}
\;\;\;\;\; 
\lambda_{ij1} =
\left( \begin{array}{cccc}
0&0&0&0\\
0&-1&0&0\\
0&0&1&0\\
0&0&0&0 
\end{array}\right) =: A^{(1)}
\nonumber
\\ 
\lambda_{ij2} 
&=&
\left( \begin{array}{cccc}
0&0&0&0\\
0&0&1&0\\
0&1&0&0\\
0&0&0&0 
\end{array}\right) =: A^{(2)}
\;\;\;\;\; 
\lambda_{ij3} 
=
\left( \begin{array}{cccc}
0&0&0&-1\\
0&0&0&0\\
0&0&0&0\\
-1&0&0&0 
\end{array}\right) =: A^{(3)}.
\nonumber
\end{eqnarray}
We denote by  $A^{(\nu),red}$ the reduced $3 \times 3$ matrices 
that appear when the first row and the first column of $A^{(\nu)}$ 
is dropped: 
$ (A^{(\nu),red})_{i,j}:=(A^{(\nu)})_{i,j=1,..,3}.$ 
In the same sense one can define a reduced tensor 
$(\lambda_{ijk}^{red})_{i,j,k}:=(\lambda_{ijk})_{i,j,k=1,..,3}.$
 
Let us now investigate what can be achieved with one 
measurement setting. One measurement setting is of the 
form
\begin{eqnarray}
M &=&  
\sum_{r,s,t=0}^{1}c_{rst}\KetBra{A_r}\otimes\KetBra{B_s}\otimes\KetBra{C_t}
\nonumber \\
&=&
\sum_{i,j,k=0}^{3}\mu_{ijk}
\sigma_{i} \otimes \sigma_{j} \otimes \sigma_{k},
\end{eqnarray}
Defining $s^A$ as the Bloch vector of $\KetBra{A_0}$ (and similarly
$s^B$ and $s^C$ for $\KetBra{B_0}$ and $\KetBra{C_0}$) and using the 
same argumentation as in the two qubit case, it is easy to see 
that the reduced $3\times 3\times 3$ tensor $\mu_{ijk}^{red}$ is 
given by
\begin{equation}
\mu_{ijk}^{red}=\sum_{r,s,t=0}^{1}c_{rst}(-1)^{r+s+t}s^A_is^B_js^C_k.
\end{equation}
Therefore for all $k$ the matrices $(\mu_{ijk}^{red})_{i,j}$ are of 
rank one. 

In order to show that $W_{GHZ}$ can not be measured with three measurement 
settings, it suffices to show that it is not possible to find 
three $3\times3$ matrices $B_i, i\in \{0,..,2\}$ of rank one such 
that $A^{(0),red}, A^{(1),red}$ and $ A^{(2),red}$ can be 
represented as linear combinations of the $B_{i}$. Let us assume 
the contrary, {\it i.e.} that we have three $B_i.$ Since the $ A^{(i),red}$
span a three dimensional subspace in the space of all $3\times3$ 
matrices, the $B_i$ have to be linear independent (as matrices) 
and have to span the same space. That would imply that 
any of the $B_i$ could be written as a linear combination of the 
$A^{(i),red}.$ But a general linear
combination of the $ A^{(i),red}$ is of the form:
\begin{equation}
\mathcal{A} =
\left( \begin{array}{ccc}
-\alpha&\beta&0\\
\beta&\alpha&0\\
0&0&\gamma 
\end{array}\right)
\end{equation}
This is of rank one if and only if $\alpha=\beta=0.$ Thus, we arrive 
at a contradiction, the $B_i$ cannot be of rank one and
linear independent.
$\hfill \Box$
\\
\\
For the investigation of W-states two witnesses were constructed in
\cite{2001:acin}. The first one is given by
\begin{equation}
W_{W1}=\frac{2}{3}\Eins-\KetBra{W},
\end{equation}
This witness detects states belonging to the W-class and the GHZ-class, 
{\it i.e.} it's expectation value is positive on all biseparable and 
fully separable states. The optimized decomposition is given by
\begin{eqnarray}
W_{W1}&=&
\frac{1}{24}
\Bigl(
17 \cdot \Eins \otimes \Eins \otimes \Eins+
7 \cdot \sigma_z \otimes \sigma_z \otimes \sigma_z+
\nonumber \\
&&
+ 3 \cdot \sigma_z \otimes \Eins \otimes \Eins +
  3 \cdot \Eins \otimes \sigma_z \otimes \Eins +
  3 \cdot \Eins \otimes \Eins \otimes \sigma_z +
\nonumber \\
&&
+ 5 \cdot \sigma_z \otimes \sigma_z \otimes \Eins + 
  5 \cdot \sigma_z \otimes \Eins \otimes \sigma_z+ 
  5 \cdot \Eins \otimes \sigma_z \otimes \sigma_z +
\nonumber\\
&&
- (\Eins+\sigma_z+\sigma_x)\otimes
  (\Eins+\sigma_z+\sigma_x)\otimes 
  (\Eins+\sigma_z+\sigma_x)
\nonumber \\
&&
- (\Eins+\sigma_z-\sigma_x)\otimes
  (\Eins+\sigma_z-\sigma_x)\otimes 
  (\Eins+\sigma_z-\sigma_x)
\nonumber\\
&&
- (\Eins+\sigma_z+\sigma_y)\otimes
  (\Eins+\sigma_z+\sigma_y)\otimes 
  (\Eins+\sigma_z+\sigma_y)
\nonumber \\
&&
- (\Eins+\sigma_z-\sigma_y)\otimes
  (\Eins+\sigma_z-\sigma_y)\otimes 
  (\Eins+\sigma_z-\sigma_y)
\Bigr).
\label{ww1zerlegung}
\end{eqnarray}
Here, only five correlated measurement settings are necessary. This 
decomposition is also optimal:
\\
\\
{\bf Proposition 3.}
The witness $W_{W1}$ can not be measured with four measurement settings, 
{\em i.e.} the decomposition (\ref{ww1zerlegung}) is optimal.
\\
\emph{Proof.} The strategy of the proof is the same as for the proof of  
Proposition 2, so we can make it short. First one computes 
$W_{W1}= 1/8 \sum_{i,j,k=0}^{3} \lambda_{ijk}\sigma_{i} \otimes 
\sigma_{j} \otimes \sigma_{k},$ and the corresponding 
$A^{(i),red}, i\in{0,..,3}.$ This time, it turns out that they 
span a four dimensional space. 

Again, it suffices to show that we cannot find 
four  matrices $B_i, i\in \{0,..,3\}$ of rank one such 
that $A^{(0),red}, A^{(1),red}, A^{(2),red}$ and 
$ A^{(3),red}$ can be represented as linear combinations 
of the $B_i.$ Here, the assumption that we have four $B_i$ 
fails due to similar reasons as above: As above, the $B_i$ 
have to be linear independent and it has to be possible
to write any of the $B_i$ as a linear combination of 
the $ A^{(i),red}.$ A general linear combination of 
the $ A^{(i),red}$ is now of the form
\begin{equation}
\mathcal{A} =
\left( \begin{array}{ccc}
\alpha&0     &\beta  \\
0     &\alpha&\gamma \\
\beta &\gamma&\delta 
\end{array}\right), 
\end{equation}
and this is of rank one if and only if $\alpha=\beta=\gamma=0,$ 
hence we arrive at a contradiction.
$\hfill\Box$
\\
\\
The second witness for W-class states is given by
\begin{equation}
W_{W2}=\frac{1}{2}\Eins-\KetBra{GHZ}.
\label{w2witness}
\end{equation}
This witness can be measured locally with the same decomposition as
(\ref{ghzwitnesszerlegung}) substracted by $\Eins/4 $. If 
$-1/4 \leq Tr(W_{W2}\varrho)\leq 0,$ $\varrho$ is threepartite 
entangled, it is either a W-state or a GHZ-state. If 
$Tr(W_{W2}\varrho)\leq -1/4,$ $\varrho$ is a GHZ-state.
It also can serve for the detection of states of the type 
$(1-p)\Eins/8+p\KetBra{W},$ this is explained in \cite{2001:acin}.

Let us conclude this section with a remark about the relationship
between convertability under SLOCC and  the number of LvNMs needed
for a local measurement.
One may interpret our results for two qubits in the following way: 
A projector $\KetBra{\phi}$ can be measured with one or three LvNMs, 
depending on whether it is a product state or not. These two classes
coincide with the two inequivalent (under SLOCC) classes for 
two qubits \cite{2000:duer}. One may think that SLOCC and LvNM  are 
in this way related. Even our results in \cite{2002:guehne2}, which 
state that the number of LvNMs needed strongly depends on the Schmidt rank of 
$\Ket{\phi}$ for $N\times N$-systems may support this conjecture, since for 
bipartite systems the Schmidt rank classifies inconvertible sets under 
SLOCC.  But our work 
on three qubits suggests that this coincidence is just by chance. 
For a general state $\Ket{\psi_{GHZ}}$ of the GHZ-class there 
always exists an orthonormal product basis  
in which it can be written as
\begin{equation}
\Ket{\psi_{GHZ}}=\lambda_0 \Ket{000}
+\lambda_1 e^{i\theta} \Ket{100}+ \lambda_2 \Ket{101}
+ \lambda_3 \Ket{110}+ \lambda_4 \Ket{111}
\end{equation}
and for a general W-state $\Ket{\psi_{W}}$ there exists the same
description, but with $\lambda_4=\theta=0$ \cite{2000:acin}. If one 
has an optimized decomposition of a general $\KetBra{\psi_{GHZ}}$
it should be possible to derive a decomposition of 
$\KetBra{\psi_{W}}$ by setting
$\lambda_4=\theta=0,$ this decomposition would need less or 
the same number of LvNMs. In the other direction, this means that 
for a general $\Ket{\psi_{W}}$  there exists a $\Ket{\psi_{GHZ}}$ 
which needs at least the same number of LvNMs for a local measurement. 
But as we have shown for $\Ket{W}$ there also exists a
GHZ-state (namely $\Ket{GHZ}$) that needs less LvNMs. 
Hence, the relation between SLOCC and LvNM seems not to be 
so simple, if there is a relation at all.

\section{Conclusion}
We have studied how three qubit entanglement can be investigated with
local measurements. For this purpose we decomposed already known 
entanglement witnesses into local measurements. We have shown that 
these decompositions are optimal. By this, we have shown that four 
measurement settings suffice for the detection of 
true threepartite entanglement and especially GHZ-type 
entanglement.

We wish to thank Dagmar Bru{\ss}, Artur Ekert, Maciej Lewenstein, 
Chiara Macchiavello and Anna Sanpera for helpful discussions. 
O.G. wishes to thank Ofer Biham and  Karol {\.Z}yczkowski for 
the discussions in Ustro{\'n}, the organizers for 
their work and last but not least for their financial support. 
This work has further been supported by the DFG (Graduiertenkolleg 282 
and Schwerpunkt ``Quanteninformationsverarbeitung'').



\end{document}